# Evolution of Variance in Offspring Number: the Effects of Population Size and Migration

Max Shpak
Department of Ecology and Evolutionary Biology, University of Tennessee.
Knoxville, TN 37996–1610 USA
(865) 974–4605,(865)974–3067(fax)
mshpak@tiem.utk.edu

## ■ Abstract

It was shown by Gillespie (1974) that if two genotypes produce the same average number of offspring on but have a different variance associated within each generation, the genotype with a lower variance will have a higher effective fitness. Specifically, the effective fitness is $w_e = w - \frac{\sigma^2}{N}$, where w is the mean fitness, $\sigma^2$ is the variance in offspring number, and N is the total population size. The model also predicts that if a strategy has a higher arithmetic mean fitness and a higher variance than the competitor, the outcome of selection will depend on the population size (with larger population sizes favoring the high variance, high mean genotype). This suggests that for metapopulations with large numbers of (relatively) small demes, a strategy with lower variance and lower mean may be favored if the migration rate is low while higher migration rates (consistent with a larger effective population size) favor the opposite strategy. Individual based simulation confirms that this is indeed the case for an island model of migration, though the effect of migration differs greatly depending on whether migration precedes or follows selection. It is noted in the appendix that while Gillespie 1974 does seem to be heuristically accurate, it is not clear that the definition of effective fitness follows from his derivation.

# ■ Introduction: Intra and Intergeneration Variance

The reproductive strategies of various organisms generally fall under two broad categories: semelparity and iteroparity. Semelparous organisms (for example, annual plants) make a single large reproductive effort, usually at the end of their lives, while iteroparous organisms (e.g. perennials) spread their reproduction out over several clutches or seasons, with no particular requirement for equal reproductive success in any season.

The evolution of semelparity and iteroparity are often addressed in terms of life history trade-offs and reproductive effort optimization (Charnov and Schaffer 1973, Schaffer 1974), but another main factor in assessing the relative successes of these strategies lies in the variance in offspring number and the intrinsic "risk spreading" and "bet hedging" nature of iteroparity (Stearns and Crandall 1981, Stearns 2000). Assuming all else is equal (namely tradeoffs between survival and reproduction are such that the semelparous and iteroparous strategies being compared give the same net reproductive outpu), it can be shown that iteroparity can neverthless remain the favored strategy because multiple reproductive strategies can be said to "spread the risk" and "hedge the player's bets." What this means intuitively is that the semelparous organism plays a strategy of "all or nothing" in its reproductive effort while the iteroparous organism, in the fashion of a gambler, staggers its risk over multiple smaller efforts.

The difference in the two strategies lie in the expected variance in surviving offspring. If a genotype i produces $k_i$ clutches of $n_i$ offspring, where each clutch survives or fails as a whole with probability $\pi_i$ (a reasonable assumption for bird's nests suffering from predation or seed crops which survive or fail as a whole due to the vagaries of rain or drought), the mean and variance in fitness are:

(1)     $w_i = \mathrm{E} x_i = n_i\, k_i\, \pi_i$

(2)     $\sigma_i^2 = \mathrm{Var}(x_i) = n_i^2\, k_i\, \pi_i\, (1 - \pi_i)$

For a semelparous organism, k=1, while for an iteroparous organims that produces the same total number of offspring over its lifetime, k>1 and the number of offspring per clutch n is less than that of the semelparous strategy. As a result, the variance of the iteroparous strategy will be much lower than for its semelparous counterpart. For example, a semelparous organism that has one clutch of 10 offspring which survive or fail with probability 0.1 has a variance $\sigma^2 = 9$, while an iteroparous organism producing 10 clutches with a single offspring over the course of its life, each with a survival probability of 0.1, has a variance of 0.9 (the mean for both is 1). Over the entire population, the effect is more pronounced when the high variance genotypes are few in number, because the variance in the sample mean is inversely proportional to the sample size (i.e. extinction due to stochastic fluctuation is more likely when the strategy with greater fluctuation is at a low initial frequency).

The higher expected net profit associated with bet–hedging has long been appreciated by gamblers and investors, who prefer to place multiple small increment bets rather than a single large one on order to gain a higher reward. The use of the geometric mean to measure the gain in expected wealth under these competing scenarios, and the subsequent demonstration that spreading one's wealth yields higher net profit, was first formally proposed by Daniel Bernoulli in 1738 and has since become a standard model for diversified portfolio building in finance (e.g. Keown et al 2001), and as an explanation for why heterogeneous assemblages of organisms tend to be more robust to environmental perturbation than homogeneous assemblages (Tilman et al 2000).

The extension of these results to evolutionary biology is quite obvious (Stearns 2000 ) because biological lineages, like investments, grow or contract geometrically rather than additively . If there is a single lineage with relatively few representatives, the effects of a low reproductive output have a much stronger effect than those generations where a large number of offspring are produced, in that a single generation of zero offspring can kill off a lineage in spite of past successes.

It is therefore desirable to have a measure of fitness which reflects the effects of the second as well as the first moment, because the arithmetic mean fitness alone is not an adequate predictor for which strategy is more likely to become fixed. A number of heuristic arguments have been made favoring the use of the geometric mean, i.e. $W_g = (\prod_{i=1}^{n} w_i)^{1/n}$, which has the desired property $W_g < w$ when $0 < \sigma^2$. There is a convenient approximation for the geometric mean in terms of arithmetic mean and variance, i.e.

$$W_g \simeq w \exp\left[-\frac{\sigma^2}{2w^2}\right] \simeq w - \frac{\sigma^2}{2w} \leq w$$

For sufficiently small w, this can be approximated as $W_g \simeq w - \frac{\sigma^2}{2}$ via a geometric expansion of $1/w^2$.

It has been shown (Haldane and Jayakar 1963, Gillespie 1973, 1977, Proulx and Day 2001) that for a stochastic environment, the geometric mean (or at least the first two terms of a Taylor expansion corresponding to the above) is an accurate measure of fitness and a predictor of fixation probability, unlike the arithmetic mean. However, internal stochasticity in number of offspring (within generation variance in fitness) is not equivalent to stochasticity in offspring survival due to environmental fluctuation (between generation variance in fitness).

In the case of offspring number variance within a generation, the selection and drift terms differ from those derived for stochastic selection. It was shown by Gillespie (1974) that the effective fitness is a function of population size N, i.e. the effective fitness of genotype i is $w_{e,i} = w_i - \frac{\sigma_i^2}{N}$. Not only is fitness decreased as a result of high variance in offspring number, but the effect is more pronounced for small populations than for large ones (which is more or less consistent with our intuition about the effects of variance, given that a lineages are more likely to become extinct due to offspring number when there are fewer of them). While certain parts of his derivation are questionable and unclear (see Appendix), this measure of effective fitness seems to predict evolutionary dynamics that are confirmed below by individual based simulations.

The effects of population size has a number of potentially interesting implications for selection on variance in metapopulations. Given two strategies, one a high mean, high variance strategy and the other a lower mean, lower variance strategy (i.e. $w_1 < w_2$, $\sigma_1^2 < \sigma_2^2$) there will be some critical population size (see Figure 1) at which the effective fitnesses are equal, so that below the critical value the lower variance strategy is more likely to become fixed, while above it the higher mean strategy is more likely to go to fixation. This critical value (which only exists when the high variance strategy is also the higher mean strategy) is $\hat{N} = \frac{\sigma_2^2 - \sigma_1^2}{w_1 - w_2}$.

## ■ Single Deme Dynamics

Gillespie (1974) derived the Kolmogorov backward equation for haploid genotypes with variance in offspring number. Collecting the terms associated with the first and second partials of allele frequency, (with $w_i = 1 + \mu_i$)

(3)
$$\frac{\partial \phi(p, N, t)}{\partial t} = p(1-p)\left(\mu_2 - \mu_1 + \frac{\sigma_1^2 - \sigma_2^2}{N}\right) \frac{\partial}{\partial p}[\phi(p, N, t)] + \frac{p(1-p)}{2N}\left((1-p)\sigma_1^2 + p\sigma_2^2\right) \frac{\partial^2}{\partial p^2}[\phi(p, N, t)]$$

(again, see the Appendix for issues relating to its derivation). The only contribution to the diffusion term is the variance in offspring number. Genetic drift due to binomial sampling of gametes is not taken into account in (3), since in the absence of offspring number variance the diffusion term in (3) is 0 (as if it were an infinite population).

It follows that the probability of fixation of a genotype with initial frequency p and fitness mean/variance $\mu_1$, $\sigma_1^2$ is

$$(4) \quad U(p) = \frac{\int_0^p ((1-x)\sigma_1^2 + x\sigma_2^2)^{2\left(\frac{N(\mu_1-\mu_2)}{\sigma_1^2-\sigma_2^2}-1\right)} dx}{\int_0^1 ((1-x)\sigma_1^2 + x\sigma_2^2)^{2\left(\frac{N(\mu_1-\mu_2)}{\sigma_1^2-\sigma_2^2}-1\right)} dx}$$

U(p) is independent of population size N when $\mu_1 = \mu_2$. As Gillespie (1974) noted, this is because the "effective fitness" contribution scales inversely with 1/N, as does the strength of selection relative to stochastic fcactors, so that these opposite tendencies cancel. Therefore, U(p) is a constant with respect to population size when the arithmetic mean fitnesses are equal. From (4) one can calculate the fixation probabilities of the various strategies for $\sigma_1^2=0.9$ versus $\sigma_1^2=9$ (these parameters correspond to $k_1=10$ clutches of a single offspring versus $k_2=1$ clutch of $n_2=10$, with survival probabilities of 0.1 per clutch) as a function of initial frequency. These values, shown for the high variance strategy in Figure 2, are the same for any population size.

In contrast, if the arithmetic means are not equal, population size does have an effect because the fitness differential does not scale as 1/N, i.e. the stochastic sample variance and the strength of selection are not precisely inversely related. If the strategy with higher variance has a higher arithmetic mean, the relative fitness values will depend on population size, with the higher variance strategy being favored in sufficiently large populations.

Consider again the case of $w_1 > w_2$ (1 vs. 0.9) and $\sigma_1^2 > \sigma_2^2$ (0.81 and 9), corresponding to competition between a genotype that produces 9 clutches with one offspring versus a single clutch of 10. The critical population size at which the two strategies are "neutral" is $\hat{N}=82$. Figure 3 plots the fixation probabilities of the strategies for different

$\hat{N}$

initial values of p. In 3a (which shows the fixation probability of the high variance, low mean strategy), both strategies have equal initial frequencies p=0.5, and it can be seen that the fixation probability is approximately 0.5 when N is just over 80.

Figure 3b plots the probability of the high variance, low mean strategy's probability of invasion (i.e. probability of fixation given an initial frequency of 1/N). The high values for very small population sizes simply reflect the artifact that the initial frequency is high for N of order unity. When the population size is sufficiently large (82<N) that the effective fitness of the high variance strategy is higher than its competitor, the invasion probability becomes higher than it was for populations of 10<N<80, but not enough to fully compensate for the effects of low initial frequency. It is interesting, however, that the fixation probability of a high variance invader is actually greater at an initial frequency of 1/N=0.001 than at 1/N=0.2 because of the effects of population size on fitness.

The last figure in the set (3c) shows the invasion probability of a low variance, low mean strategy from an initial frequency of 1/N. As expected, the invasion probabilities for small N are high due to two factors: a high initial frequency, and a high effective fitness at low population sizes.

■ **Individual Based Simulations: Selection on Variance in Offspring Number in a Single Deme**

The analytical results in the previous section and in Gillespie (1974) are qualitatively consistent with the outcome of individual based simulations for selection on mean offspring number and variance. The simulations were written in the C programming language and copies of the program code are available from the author upon request.

In every simulation, a certain proportion of individuals p in a population of N are chosen to be semelparous, the rest are iteroparous. During the simulated life cycle,

$n_1$                                                                  $k_2$

every individual produces either on average $n_1$ offspring or an average of $k_2$ clutches (in both cases chosen from a Poisson distribution) of single offspring, which survive or fail as clutches with probabilities $\pi$, $1-\pi$. Those remaining are pooled into an offspring population, and a sample of N offspring is chosen for the next generation. Selection is "soft," (i.e. Levene 1953, Wallace 1968) and population size remains constant across iterations. Generations are non-overlapping, so that the only contributions to the population at time t+1 are the offspring of individuals at time t.

    Two sets of simulations were conducted. The first set had equal initial frequencies for both strategies, while in the second set, the entire population (except for one individual) was set to a given strategy in order to investigate the probability of invasion of high and low variance genotypes. The simulations were run for 1000 generations, wsufficiently many so that one strategy is almost always fixed in the population (because selection is effectively directional or neutral in spite of fluctuating relative fitnesses and there is not true frequency dependence, stable coexistence can probably be excluded) . In turn, there were 1000 runs of each 1000 generation cycle so that fixation probabilties could be averaged over multiple runs (note that in all simulations 1000 generations was enough time for one allele to become lost or fixed by the end of each run, so that transient frequencies never entered into the estimations).

    Figures 4a,b plot the fixation probabilities given equal initial frequencies of two strategies as a function of population size. In 4a, the strategies have equal arithmetic mean fitness ($w_1 = w_2 = 1$) but different variances ($\sigma_1^2 = 0.9$, $\sigma_2^2 = 9$). The analytical solutions to the diffusion equations predict that the fixation probability of either strategy is independent of population size. For an initial frequency of 0.5, the low variance strategy has a fixation probability of approximately 0.8 (Figure 4a) for a wide range of population sizes, which qualitatively is quite close to the analytical prediction of U(0.5)=0.82.

$w_1$

$w_2$

The only anomaly is the slightly higher fixation probabilities for very small population sizes (N<50), which are probably due to the effects of genetic drift in the simulations.

In 4b, the high variance strategy has a higher arithmetic mean, i.e. ($w_1$=0.9, $w_2$=1) with corresponding variances calculated from the number of clutches and survival probabilities ($\sigma_1^2$=0.81, $\sigma_2^2$=9). In a small population the lower variance strategy has a slightly higher fixation probability given equal initial frequencies, as the effective fitness of the low variance strategy is higher in spite of its lower arithmetic mean, while for sufficiently large population sizes (N at 100 or greater) the higher variance, high mean strategy has a higher effective fitness and probability of fixation. Qualitatively, it is similar to the analytical predictions for the same parameters shown in 3a.

The probability of invasion by a high variance (see Figure 4c) strategy is invariably low. Unless $w_1 \ll w_2$ the fixation probabilities of a mutant with higher variance than the resident genotype will always be very low. The higher variance strategy has to reach a relatively high frequency in order for its effective fitness to be higher than the resident, and the only way of doing so is through drift running counter to initial negative selection.    The match between 4c and the corresponding analytical prediction of fixation probability (Fig. 3b) is quite poor for small populations, presumably due to the high probability of losing an unfavorable rare allele due to both genetic drift and selection. Since the results of every set of simulations were averaged over multiple (1000) trials, the difference is not due to sampling error alone, but to the fact that the diffusion equations and their solutions (Eqs. 3–4) do not includ the effects of genetic drift proper, which is a factor in individual based simulations as it is in nature. This also accounts for the non−constant fixation probabilities of the low variance strategy in the case where arithmetic means are equal (i.e. Figure 4a).

In contrast, the analytical predictions for invasion probabilities of the low vari-ance strategy in 3c are a fairly good match to the individual based simulation results in 4d, presumably again due to the combined effects of high frequency and effective fitness outweighing sampling error in the simulations.

Some of the discrepancy between individual based simulations and analytical results is also due to the fact that the diffusion approximation only gives accurate predictions of selection and sampling dynamics under a restrictive range of parameters. In the case of selection for variance in offspring number, the coefficients associated with the diffusion and drift terms alike are quite high due to the large variances in the number of progeny for one of the strategies. If selection coefficients are of higher order than the variance contributions, the assumptions behind the approximation start to break down because the higher moments associated with the selection term become significant. These limitations of the diffusion approximation are discussed in Kimura (1964) and in Ewens (2003).

## ■ Migration: Effective Population Size in a Metapopulation

Because the effective fitness (4) of a strategy depends on population size, factors that influence the number of individuals of a given genotype sampled in each generation can alter the likelihood that a given strategy will become fixed or lost.

In particular, consider a metapopulation (Levins 1968, Hanski and Gilpin 1991) consisting of D demes, each with N individuals. If there is no migration in the system, then clearly the dynamics are determined by the number of individuals N within each independent deme. If on the other hand migration rates are high enough so that an individual in any given deme is just as likely to have had a parent in another deme as in its current place of residence, one would expect that the genotypes favored in a population of size DN to have the higher probability of fixation. For intermediate migration rates, the evolutionary dynamics should be reflective of an effective population size of somewhere between N and DN . In particular, for a large number of small demes, we would expect that a high variance, high mean strategy may be favored given sufficiently high migration rates, while if migration is low the low variance strategy appropriate to a smaller population is more likely to become fixed.

The exact effect of migration will also depend on the organism's life cycle (i.e. the order in which reproduction, migration and selection take place). Consider first a life cyclewhere reproduction followed by (soft) selection occurs within each deme, and only afterwards is there migration where every deme exchanges a fraction m with each of the (D−1) remaining demes. In this case, the relevant effects of population size all take place within the small demes, and the only metapopulation effect is one of averaging over demes by "mixing" after selection. It is predicted in this case that the effective population size and the measured fitness of each strategy should not differ substantially from a model with no migration, apart from the effects of inhomogeneity of allele frequencies between demes.

Note that here the term "effective population size" is used in a somewhat different context than its conventional use in population genetics theory. Normally, effective size is defined with respect to the process of genetic drift, as the size the population would be in the absence of subdivision, unequal sex ratios, etc. to give the same probability of loss or fixation of neutral alleles due to drift. Here, the "effective population size" is used with respect to the process of selection. Because the effective fitness of a strategy depends on population size and its frequency, we define "effective population size" as the value $N_e$ that would give a strategy with a particular variance the same effective fitness in a non−structured population. Consequently, the effective size will be with respect to a given strategy, and thus dependent on both the strategy's frequency in differ-ent demes and the total number of individuals.

The "effective population size" and effective selection coefficient under an island population model (e.g. Wright 1931, Kimura 1953) without spatial structure are derived from the diffusion approximation, where the parameter m is the proportion of individuals that the Ith deme exchanges with any one of its D−1 neighbors (therefore, a proportion m(D−1) of each deme's offspring are found in another deme in the next generation, with $x_{J_i}$ denoting the frequency of the ith allele in the Jth deme). We follow Gillespie's example of deriving the forward equation of allele frequency from the multi-variable diffusion equation on absolute frequencies in a diallelic system:

(5)
$$\frac{\partial \phi(x_{1_1} \ldots x_{D_1}, x_{1_2} \ldots x_{D_2}, t)}{\partial t} =$$
$$\sum_J \sum_j \frac{\partial}{\partial x_{J_j}} \left[ -\mu_j \left( (1-(D-1)m) x_{J_j} + m \sum_{K \neq J} x_{K_j} \right) \phi(x,t) \right] +$$
$$\frac{1}{2} \frac{\partial^2}{\partial x_{J_j}^2} \left[ \sigma_j^2 \left( (1-(D-1)m_j) x_{J_j} + m \sum_{K \neq J} x_{K_j} \right) \phi(x,t) \right]$$

performing a change of variables in a diallelic case and collecting terms associated with the first derivate with respect to $p_I$ (written as p below by an abuse of notation), with $x_1$ and $x_2$ the absolute frequencies of alleles in the Ith deme, the differential operators are:

(6)
$$p = \frac{x_1}{x_1 + x_2}, \quad N = x_1 + x_2$$
$$\frac{\partial \phi}{\partial x_1} = \frac{\partial \phi}{\partial p} \frac{\partial p}{\partial x_1} + \frac{\partial \phi}{\partial N} \frac{\partial n}{\partial x_1} = \frac{1-p}{N} \frac{\partial \phi}{\partial p} + \frac{\partial \phi}{\partial N};$$
$$\frac{\partial \phi}{\partial x_2} = \frac{-p}{N} \frac{\partial \phi}{\partial p} + \frac{\partial \phi}{\partial N};$$
$$\frac{\partial^2 \phi}{\partial x_1^2} = \left(\frac{1-p}{N}\right)^2 \frac{\partial^2 \phi}{\partial p^2} + \frac{2(1-p)}{N} \frac{\partial^2 \phi}{\partial N \partial p} + \frac{\partial^2 \phi}{\partial N^2};$$
$$\frac{\partial^2 \phi}{\partial x_2^2} = \left(\frac{-p}{N}\right)^2 \frac{\partial^2 \phi}{\partial p^2} - \frac{2p}{N} \frac{\partial^2 \phi}{\partial N \partial p} + \frac{\partial^2 \phi}{\partial N^2};$$

Following Gillespie in setting all derivatives with respect to N to zero (assuming a constant population size), the Kolmogorov forward equation for the distribution of $p_I$ is:

(7)
$$\frac{\partial \phi[p_1 \ldots p_D, t]}{\partial t} =$$
$$(1 - (D-1)m)(-\mu_1 - \mu_2)\phi[p_1 \ldots p_D, t] -$$
$$\left[(1-p_I)\left[\mu_1\left(((1-(D-1)m)p_I) + m\sum_{K \neq I} p_K\right) - \frac{\sigma_1^2}{N}\left(1 - m\sum_{K \neq I}(1-p_K)\right)\right] +\right.$$
$$p_I\left[\mu_2\left((1-(D-1)m)(1-p_I) + m\sum_{K \neq I}(1-p_K)\right) -\right.$$
$$\left.\left.\frac{\sigma_2^2}{N}\left(1 - m\sum_{K \neq I} p_K\right)\right]\right] \frac{\partial \phi[p_1 \ldots p_D, t]}{\partial p_I} +$$
$$\frac{1}{2N}\left(\sigma_1^2(1-p_I)^2\left[(1-(D-1)m)p_I + m\sum_{K \neq I} p_K\right] -\right.$$
$$\left.\sigma_2^2 p_I^2\left[(1-(D-1)m)(1-p_I) + m\sum_{K \neq I}(1-p_K)\right]\right)\frac{\partial^2 \phi[p_1 \ldots p_D, t]}{\partial^2 p_I}$$

the selection coefficient with respect to the frequency in the Ith deme is (the caveats for coefficients associated with the variance terms are discussed in the Appendix):

(8)
$$M(p_I) =$$
$$(1-p_I)\left[\mu_1\left((1-(D-1)m)p_I\right) + m\sum_{K \neq I} p_K - \frac{\sigma_1^2}{N}\left((1-(D-1)m)p_I + m\sum_{K \neq I} p_K\right)\right] -$$
$$p_I\left[\mu_2\left((1-(D-1)m)(1-p_I) + m\sum_{K \neq I}(1-p_K)\right) -\right.$$
$$\left.\frac{\sigma_2^2}{N}\left((1-(D-1)m)(1-p_I) + m\sum_{K \neq I}(1-p_K)\right)\right]$$

The "effective population size" $N_{e,I}$ for a single deme with respect to the fitness consequences of N are evaluated by comparing the selection differentials for demes in a metapopulation to a single deme of comparable size, i.e.

$$\frac{\sigma_1^2 \, p_I \, (1 - p_I)}{N_{e,I}} =$$

$$\frac{\sigma_1^2}{N} (1 - p_I) \left[ (1 - (D-1) m) \, p_I + m \sum_{K \neq I} p_K \right],$$

$$(9) \quad N_{e,I} = \frac{N p_I}{((1 - (D-1) m) \, p_I + m \sum_{K \neq I} p_K)}$$

For the metapopulation as a whole, it is proposed that the effective population size will simply be the average across all of the demes,

$$N_e = \frac{\sum_I N_{e,I}}{D}$$

It is clear that when m=0, the effective population size will be $N_e$=N. Furthermore, if the allele frequencies are the same in every deme, $\hat{N}$=N irrespective of the migration rate. Only if there is asymmetry in allele frequencies in different demes does the effective population size differ from the census number (as defined by fitness effects due to offspring variance). If the average frequency in the metapopulation is some value $\bar{p}$, with the Ith deme having $p_I < \bar{p}$, $N_{e,I}$ <N for a nonzero migration rate. The reverse is true for a deme with a higher frequency. Figure 5a plots effective population size for a range of frequencies and a migration rate of 0.01 against variance in $p_I$, constrained by mean metapopulation frequency of 0.5. The same is shown for a migration rate of 0.05 in 5b.

Note the asymmetry here: while the high frequency $p_J$ deme has $N_{e,J}$ somewhat larger than N, the extent to which the 1–p frequency deme has a diminished $N_{e,I}$ is greater in magnitude. For example, if the metapopulation with D=10 demes has mean frequency $\bar{p}$=0.5, a deme with a frequency 0.1 has an effective population size with respect to selection for variance of $N_e$=41.7 when m=0.01 and $N_e$=16.7 when m=0.1. When the deme frequency is 0.9, the respective effective population sizes are $N_e$=51.1 and 64.3. In any metapopulation mean frequency $\bar{p}$, this model of migration will lead to an overall decrease in effective population size in all of the demes with increasing migration rate and increasing asymmetry in allele frequencies between demes.

Unless there are great asymmetries in allele frequencies in different demes, the difference between the migration model outlined above and one without migration should not be pronounced. Indeed, individual based simulations on 10 demes of 50 individuals behave much like single N=50 populations when the initial frequencies are set to 0.5 or 1/50 in every deme (favoring the low variance strategy even when the mean fitness of the high variance strategy is somewhat greater, as in Figures 3 and 4). The only noticeable effect is that for higher migration rates the fixation probability of the high variance strategy becomes higher (almost always at unity for nonzero values of m) compared to values of 0.98 (for p=0.5) and 0.94 (for p=0.02). This reflects the lower average effective population size caused by inhomogeneities in allele frequency across demes, which is likely to arise as a consequence of genetic drift in each trial. The effects are even more prounounced if one begins with an asymmetry where half the demes are near fixation for one strategy and the rest at near fixation for the other.

If the sequence of events in the life cycle is reversed, i.e. reproduction and migration occur prior to selection, the dynamics and influence of population size are entirely different. Consider a set of D demes where migration between demes occurs prior to reproduction but before selection. Where in the previous case all selection occured within a deme of size N and the only effect of migration was due to differences in intrademic allele frequency, here the actual pool of individuals that can contribute to a deme prior to selection is larger due to migration. Because the fitness decrement due to offspring variance varies inversely with the number of individuals sampled in each generation, migration prior to selection should, all else being equal, decrease the effects of variance in offspring number.

A heuristic for the effective size of the metapopulation as a whole is the size of the pool from which an individual from any deme could come from. In the absence of migration, there are only N such choices, while with full mixing (m=1/D), the effective pool is the full metapopulation size ND. For an intermediate migration rates, the number of individuals contributing the the "migrant pool" is mDN, while in every deme there are a remaining (1−m)DN individuals. It is proposed without proof that the effective population size of the metapopulation is the weighted average of the non−migrants and the migrant pool, i.e.

(10)    $N_e = (1 - m) N + \text{mDN}$

so that allele frequencies in the metapopulation should behave as if there were a single deme of size $N_e$. This proposed estimate of effective population size is proposed without proof, as there doesn't seem to be a non−circular means of deriving it directly from the diffusion equations (i.e. without assuming a higher effective size in deriving the diffusion equations).

When migration is near zero, (10) predicts that in a case where there are 10 demes of 50 individuals each and two competing strategies where the high variance strategy is also has a higher arithmetic mean fitness (i.e. $w_1$=0.9, $w_2$=1, $\sigma_1^2$=0.81, $\sigma_2^2$=9) the effective population size is near 50 and the low variance strategy should tend towards fixation. In contrast, when m is sufficiently high for $N_e$=82 (the critical value for the effective fitnesses to be equal in a single deme). Specifically, the migration rate which produces an effective population size $N_e$ is m=$\frac{N-N_e}{N(1-D)}$, which for the critical value in this example is m=0.07 (corresponding to an average of just over 3.5 total migrants from each deme every generation).

Individual based simulations confirm these heuristic results. All simulations are for D=10 demes, each with N=50 individuals. Starting with initial frequencies at 0.5 and 0.02 (1 invading genotype of a given strategy in each deme) for both the high and low variance, Figure 6a,c plots the probability of fixing the low variance strategy as a function of the number of migrants exchanged between individual demes for p=0.5 and p=0.02, while 6b does the same for the high variance, high mean strategy.

It can be seen that with low migration, the low variance strategy has a much higher probability of fixation , while for higher migration rates, the higher variance, higher mean strategy starts to enjoy an advantage, as it would in a single deme of a larger population size. Even for the relatively low migration rate of 1.5 (corresponding to a total of 13.55 migrants per deme), the high mean, high variance strategy has a higher probability of fixation given equal initial frequency.

The migration rate at which the effective fitness of the high variance strategy becomes higher than that of the low variance genotype is somewhat higher than the value predicted from (10). Whether this is due to genetic drift in the simulations or to the fact that the (10) is not the actual effective population size for this model of migration is unclear.

Furthermore, in any model combining selection and migration, there is an additional caveat in that the diffusion approximation assumes that selection, reproduction, and migration occur more or less simultaneously. The individual based simulations suggest that the order in which migration and selection occur do in fact matter, consequently, the diffusion approximation seems to correctly predict the behavior of the process when selection occurs prior to migration but not the reverse.

## ■ Discussion: Variance and Bet−Hedging

The results for competition between high and low variance strategies found here are qualitatively concordant with the work of others. For equal mean numbers of off-spring, the higher variance strategy will tend to be disfavored for reasons outlined in the introduction, while in a high variance strategy with a higher arithmetic mean there is a trade-off between gain in "effective fitness" due to a higher mean versus a cost to having a higher variance in offspring. One can readily imagine scenarios where such a trade-off exists in nature, namely, organisms can produce more offspring, but in doing so, there is a higher probability of clutch failure due to limited resources.

To use a concrete example touched upon in the introduction, there may be a trade-off between semelparity and iteroparity, where semelparity allows a larger total reproductive output while iteroparity gives a lower variance in surviving offspring. The extent of the trade-off itself depends on parameters such as population size and initial frequencies of the strategies in question. Consequently, in competitions between iteroparous and semelparous strategies in nature, the probability of fixation of one or the other genotype will not be determined by mean and variance alone. This suggests that any empirical studies of the evolution of iteroparity, semelparity, or other changes in offspring variance should take into account the implicit frequency and density dependence of the process.

The results for multideme models with migration suggest that metapopulation dynamics may further complicate selection for high or low variance strategies. If a metapopulation consists of many small demes, selection may favor a strategy with lower variance and lower mean locally while favoring the opposite strategy "globally" given sufficient migration. It would seem that for a low or intermediate migration rate a fast/slow dynamic could arise where short-term quasi-equilibria in favor of the low variance, low mean strategy occur within each deme, while a long term dynamic drives the high variance, high mean strategy to fixation. Because the complexity of the model with multiple demes, neither the existence nor non-existence of such behavior could be proven, but at least for the parameters investigated with the individual based simulations, no fast/slow dynamics were observed. A strategy that was initially favored by selection would tend to remain favored throughout the process, and any deviation from this pattern could readily be attributed to drift. Furthermore, the calculated "effective fitness" for a migration model does not imply any kind of time dependence. A strategy is either more or less fit given the parameters related to reproduction and the migration rate.

Finally, it is worth making some general remarks about the evolution of variance in offspring number in the broader context of biological bet–hedging. By producing multiple clutches with fewer offspring, organisms can reduce the variance in fitness by spreading the risk. This applies to both within–generation variance in offspring number (treated here and in Gillespie 1974) and to variance between generations due to a fluctuating environment (Gillespie 1973, Ewens 2003). While the estimated quantity to be optimized differs in the two cases ($w_e = w - \frac{\sigma^2}{N}$ for within–generation variance, $w - \frac{\sigma^2}{2w}$ or the geometric mean in the case of a varying environment), both increase effective fitness by decreasing variance.

Risk–spreading may have wider implications in evolutionary biology. In particular, sexual reproduction and genetic recombination may be seen as one means by which organisms hedge their bets to deal with fluctuating environments within or between generations (Maynard Smith 1975). A phenotypically diverse set of offspring will be more likely to have some individuals who are suited to particular environmental conditions. If the geometric mean is indeed a good measure for the long–term performance of an evolutionary strategy, then one would expect that sexual reproduction, by producing a range of phenotypes in each generation, would have the higher expected fitness over several generations.    The higher geometric mean and lower variance in surviving offspring associated with sexual reproduction was proposed by Doebeli and Koella (2001) to stabilize population size fluctuations in much the same way that iteroparity and lower offspring variance has been shown to stabilize population size (Luethy 2000, Koella 2001 unpublished). If there is hard selection and populations can expand or contract in every generation (which further complicates matters in that relative fitnesses change with population size), then a strategy that leads to wide fluctuations can potentially drive the population to extinction even when the mean growth rate is higher than that of a competing low variance strategy. The long–term stability of populations imposes a population–level selection process that can potentially run counter to individual level selection. Consequently, even if individual selection favors a high mean and high variance strategy, in the long term the high variance strategy may still go to extinction because demes where the strategy is fixed are more likely to crash. As a result, group selection (Wilson 1983) may favor a low variance "bet hedging" strategy even where the opposing strategy is favored by individual selection.

In summary, the evolution of variance in reproductive success may be behind a range of phenomena in life history evolution, demograhics, and possibly even the origin of genetic systems and macroevolutionary trends. It has been shown that even in a relatively simple system: variance in offspring number combined with interdemic migration, the population and evolutionary dynamics will be quite different than in the case of selection on mean numbers of offspring alone.

## ∎ Acknowledgements

of the individual based simulations for a single deme conducted in this study. The author also wishes to thank the following people for advice on diffusion equation applications: Aaron King, Warren Ewens, and Sergey Gavrilets, as well as Michael Doebeli, Michael Kopp, Takehiko Hayashi, and Steve Ellner for general discussion. This work was supported by NSF grant DEB−0111613 and NIH grant GM56693 to Sergey Gavrilets.

# ■ Appendix: Some Questions and Comments On the Derivation of Effective Fitness and Population Size in Gillespie (1974)

The measure of effective fitness in Gillespie 1974 (G74) for within generation variance in offspring number was used in this study because at least as an approximation it correctly predicts the fixation probabilities of various strategies in individual based simulations, and because it offers a useful heuristic for predicting the effect of migration (and effective population size) on the relative success of strategies in a metapopulation.

However, a straightforward analysis by substituting differential operators (following his own procedure) does not give a result consistent with G74. Following his method, we begin with a bivariate diffusion equation on the absolute numbers of two haploid genotypes (Feller 1951) and performing a change of variables by expressing the absolute numbers in terms of population size n and allele (genotype) frequency p.

If for two genotypes the mean numbers of offspring are $w_1=1+\mu_1$ and $w_2=1+\mu_2$ and the variances in offspring number produced in each generation are $\sigma_1^2$ and $\sigma_2^2$, the bivariate Kolmogorov forward equation is:

(A.1)
$$\frac{\partial \phi(x_1, x_2, t)}{\partial t} = -\sum_i \mu_i \frac{\partial}{\partial x}[x_i \phi(x)] + \frac{\sigma_i^2}{2} \frac{\partial^2}{\partial x^2}[x_i \phi(x)]$$

where $x_0=x(0)$, as the backward equation describes the probability distribution on initial frequencies given a frequency at time t while the forward equation describes the distribution of frequencies at time t given an initial frequency (Kimura 1964). Note that in this model, the only contribution to the diffusion term is due to the variance in the number of offspring, gametic sample variance (genetic drift proper) is not taken into account in this where $x_0 = x(0)$, as the backward equation describes the probability distribution on initial frequencies given a frequency at time t while the forward equation describes the distribution of frequencies at time t given an initial frequency (Kimura 1964). Note that in this model, the only contribution to the diffusion term is due to the variance in the number of offspring, gametic sample variance (genetic drift proper) is not taken into account in this model for the sake of simplicy.

In order to derive the terms for expected directional change and variance of allele frequencies (see Eq. 6 in the main text), the forward equation in terms of p and N with the differential terms of $\phi$ grouped together is:

(A.2)

$$\frac{\partial \phi(p,t)}{\partial t} = (-\mu_1 - \mu_2)\phi + \left[p(1-p)(\mu_2 - \mu_1) + \frac{((1-p)\sigma_1^2 - p\sigma_2^2)}{N}\right]\frac{\partial \phi}{\partial p} - \left(\frac{p(1-2p-p^2)\sigma_1^2 + p^2(1-p)\sigma_2^2}{2N}\right)\frac{\partial^2 \phi}{\partial p^2} + (\sigma_1^2 + \sigma_2^2 - N(\mu_1 p + \mu_2(1-p)))\frac{\partial \phi}{\partial N} + p(1-p)(\sigma_1^2 - \sigma_2^2)\frac{\partial^2 \phi}{\partial N \partial p} + \left(\frac{N}{2}(p\sigma_1^2 + (1-p)\sigma_2^2)\right)\frac{\partial^2 \phi}{\partial N^2}$$

If it is assumed that population size is constant, then the distribution $\phi(p,N,t)$ has a local optimum at $N=\hat{N}$, and all of the partial derivatives with respect to N can be set to 0.

The desired structure of a Kolmogorov forward equation on p is then of the form:

(A.3)

$$\frac{\partial \phi(p,t)}{\partial t} = -\frac{\partial}{\partial p}[M(p)\phi(x)] + \frac{1}{2}\frac{\partial^2}{\partial p^2}[V(p)\phi(x)]$$

V(p) is obtained in a straightforward manner from A.2. The term $\frac{\partial^2 \phi}{\partial p^2}$ in A.2 will (from the sum rule of derivatives) have the coefficient V(p). Therefore:

(A.4 a)    $V(p) = \frac{p(1-p)}{N}((1-p)\sigma_1^2 + p\sigma_2^2)$

To obtain M(p), we note that the coefficient associated with $\frac{\partial \phi}{\partial p}$ in (A.2) has two contributions from (A.4): $C\frac{\partial \phi}{\partial p} = (\frac{\partial V(p)}{\partial p} - M(p))\frac{\partial \phi}{\partial p}$. If the partial derivatives with respect to n are included, the derivatives with respect to n of the coefficient associated with $\frac{\partial^2 \phi}{\partial N \partial p}$ also contribute to the right-hand side, here they are zero regardless).

Solving for M(p), we obtain

To obtain M(p), we note that the coefficient associated with $\frac{\partial \phi}{\partial p}$ in (A.2) has two contributions from (A.4): $C \frac{\partial \phi}{\partial p} = (\frac{\partial V(p)}{\partial p} - M(p)) \frac{\partial \phi}{\partial p}$. If the partial derivatives with respect to n are included, the derivatives with respect to n of the coefficient associated with $\frac{\partial^2 \phi}{\partial N \partial p}$ also contribute to the right–hand side, here they are zero regardless).

Solving for M(p), we obtain

$$(A.4b) \quad M(p) = p(1-p)\left(\mu_1 - \mu_2 - \frac{3}{N}(\sigma_1^2 - \sigma_2^2)\right)$$

Which differs from Gillespie's result in having an additional factor of 3 associated with the variance terms (Which, if correct, predicts a qualitatively different effective fitness. For example, the critical population size where the fitnesses are equal for a system with $\mu_1 > \mu_2$ and $\sigma_1^2 > \sigma_2^2$ is 3 times larger than it is with Gillespie's measure). In spite of the inconsistencies in his derivation, G74 is more consistent with the numerical results, hence its use throughout the paper in derivations involving migration terms.

Furthermore, the coefficients of the $\phi$ term in A.2 differ from what is derived from G74 eq. 3, which is

$$(A.5)$$

$$\frac{\partial \phi(p, N, t)}{\partial t} = -\frac{\partial}{\partial p}\left[p(1-p)\left(\mu_1 - \mu_2 - \frac{\sigma_1^2 - \sigma_2^2}{N}\right)\phi(p, N, t)\right] +$$
$$\frac{1}{2}\frac{\partial^2}{\partial p^2}\left[\frac{p(1-p)}{N}((1-p)\sigma_1^2 + p\sigma_2^2)\phi(p, N, t)\right] -$$
$$\frac{\partial}{\partial N}[N(\mu_1 p + \mu_2(1-p))\phi(p, N, t)] + \frac{1}{2}\frac{\partial^2}{\partial N^2}[N(p\sigma_1^2 + (1-p)\sigma_2^2)\phi(p, N, t)] +$$
$$\frac{\partial^2}{\partial N \partial p}[p(1-p)(\sigma_1^2 - \sigma_2^2)\phi(p, N, t)]$$

calculating derivatives and collecting coefficients here, one obtains the following terms which differ from those in A2 (the coefficients for the first derivative with respect to N also differ, but are not shown as they do not influence the outcome):

$$(A.5a) \quad -((\mu_1 + \sigma_1^2)(1-p) + (\mu_2 + \sigma_2^2)p)\phi$$

$$(A.5b) \quad \left(p(1-p)(\mu_2 - \mu_1) + \frac{1-2p}{N}((1-p)\sigma_1^2 - p\sigma_2^2)\right)\frac{\partial \phi}{\partial p}$$

In summary, at least using Gillespie's change of variables applied to the bivariate diffusion equation, the differential terms in the expansion are not consistent with G74

$$M(p) = p(1-p)\left(\mu_1 - \mu_2 + \frac{\sigma_1^2 - \sigma_2^2}{N}\right)$$

It should be noted in passing that the outcome of the change of variables depends entirely on the choice of differential operators. As far as the author understood Gillespie's methods, his change of variables were followed. However, if one uses a parameterization that assumes n fixed from the start, i.e.

$$p = \frac{x_1}{N}, \quad N = x_1 + x_2$$
$$\frac{\partial \phi}{\partial x_1} = \frac{\partial \phi}{\partial p}\frac{\partial p}{\partial x_1} = \frac{1}{N}\frac{\partial \phi}{\partial p}; \quad \frac{\partial \phi}{\partial x_2} = \frac{-1}{N}\frac{\partial \phi}{\partial p};$$
$$\frac{\partial^2 \phi}{\partial x_1^2} = \left(\frac{1}{N}\right)^2 \frac{\partial^2 \phi}{\partial p^2}; \quad \frac{\partial^2 \phi}{\partial x_2^2} = \left(\frac{-1}{N}\right)^2 \frac{\partial^2 \phi}{\partial p^2};$$

and substitute these differential operators into (A.1), we obtain and entirely different diffusion equations for frequency p:

$$\frac{\partial \phi(p,t)}{\partial t} = (\mu_2 - \mu_1)\phi + ((1-p)\mu_2 - p\mu_1 - \sigma_2^2 + \sigma_1^2)\frac{\partial \phi}{\partial p} + \frac{1}{2N}(p\sigma_1^2 + (1-p)\sigma_2^2)\frac{\partial^2 \phi}{\partial p^2}$$

which, while technically correct, corresponds to neither the previous calculations nor to Gillespie's result. It is therefore possible that Gillespie's result may follow from the correct choice in a change of variable (in any case, his results are consistent with the results of individual based simulations in predicting "critical" N while the above results are not, which is why they were used in calculating effective population size with migration), but the author was unable to reconstruct this result following the proposed change of variables in G74.

A number of technical problems can arise in transforming a bivariate equation to a univariate equation, so that the derived univariate diffusion equation depends on the change of variables, any number of which may be consistant with a constant N. However, an investigation of the problem is beyond the scope of this paper.

## ■ References

Bernoulli, D. (1738, translated and reprinted 1954). Exposition of a new theory on the measurement of risk. Econometrica 22: 23–36

Charnov, E.L. and W.M. Schaffer (1973). Life history consequences of natural selection: Cole's result revisited. American Naturalist 107: 791–793

Doebeli, M. and J. Koella (2001). Sex and population dynamics. Proceedings of the Royal Society of London B 257:17–23

Ewens, W.J. (2003). Mathematical population genetics. Springer–Verlag, New York

Feller, W. (1951). Diffusion processes in genetics. Proceeding of the 2d Berkeley Symposium on Mathematical Statistics and Probability 227–246

Gillespie, J.H. (1974). Natural selection for within–generation variance in offspring number. Genetics 76: 601–606

Gillespie, J.H. (1974). Polymorphism in patchy environments. American Naturalist 108:145–151

Gillespie, J.H. (1977). Natural selection for variance in offspring numbers: a new evolutionary principle. American Naturalist 111: 1010–1014

Haldane, J.B.S. and S.D. Jayakar (1963). Polymorphism due to selection of varying deirections. Journal of Genetics 58: 237–242

Hanski I. and M. Gilpin 1991. Metapopulation dynamics: brief history and conceptual domain. Biological Journal of the Linnean Society 42: 3–16

Keown, A.J., J.W. Martin, W.D. Petty, and D.F. Scott (2001). Financial management: principles and applications (9th edition). Prentice Hall, New York.

Kimura, M. (1953). Stepping stone model of populations. Annual Report of the National Institue of Genetics 3: 62–63

Kimura, M. (1964). The diffusion model in population genetics. Journal of Applied Probability 1:177–232

Levene, H. (1953). Genetic equilibrium when more than one niche is available. American Naturalist 87: 311–313

Levins, R. 1970. Extinction. pgs. 77–107. In M. Gestenhaber (editor). Some mathematical problems in biology. American Mathematical Society, Providence, RI.

Luethy, R. 2000. Population dynamics and the evolution of bet–hedging strategies in stochastic environments (unpublished).

Maynard Smith, J. (1975). The evolution of sex. Cambridge University Press, Cambridge UK

Proulx, S.R. and T. Day (2001). What can invasion analysis tell us about evolution in stochastic populations? Selection 2: 1–15

Schaffer, W.M. (1974). Optimal reproductive effort in fluctuating environments. American Naturalist 108: 783–790

Stearns, S.C. (2000). Daniel Bernoulli (1783): evolution and economics under risk. Journal of Bioscience 25:221–228

Stearns, S.C. and R.E. Crandall (1981). Bet–hedging and persistence as adaptations for colonizers. In Evolution today (eds. G.G.E. Scudder and J.L. Reveal). Hunt Institute, Philadelphia, PA

Wallace, B. (1968). Topics in population genetics. W.W. Norton and Company, New York

Wilson, D.S. (1983). The group selection controversy: history and current status. Annual Review of Ecology and Systematics. 14: 159–187

Wright, S. (1949). Population structure and evolution. Proceedings of the American Philosophical Society. 93:471–478

# ■ Figure Captions

**Figure 1**:
Effective fitness as a function of population size
a. The mean fitness values are equal ($w_1=w_2=1$) while $\sigma_1^2=0.9$ and $\sigma_2^2=9$. The effective fitness is here shown as a function of population size (the low variance strategy is always more fit, though the fitness values converge asymptotically at large N).
b. As above, only the high variance strategy also has a higher mean fitness ($w_2=1$

**Figure 1**:
Effective fitness as a function of population size
a. The mean fitness values are equal ($w_1=w_2=1$) while $\sigma_1^2=0.9$ and $\sigma_2^2=9$. The effective fitness is here shown as a function of population size (the low variance strategy is always more fit, though the fitness values converge asymptotically at large N).
b. As above, only the high variance strategy also has a higher mean fitness ($w_2=1$, $w_1=0.9$).

**Figure 2**:
Fixation probabilities of the high variance strategy are calculated from the Kolmogorov backward equations using numerical integration, here shown as a function of initial allele frequency. The arithmetic mean fitnesses are equal and the fixation probability depends only on initial frequency and the variance, not on population size. The fixation probability of the higher variance strategy (for $\sigma_1^2=9$, $\sigma_2^2=0.9$) is shown as a function of initial frequency.

**Figure 3**:
Fixation probabilities are calculated for various population sizes from the Kolmogorov backward equations using numerical integration.
a. In the case where the genotype with a higher mean fitness ($w_1=1$, $w_2=0.9$) also has a higher variance ($\sigma_1^2=9$ and $\sigma_2^2=0.81$), the fixation probability of the high variance strategy is calculated given an initial frequency p=0.5 for a range of population sizes.
b. Here, the invasion probability of the high variance strategy is calculated (i.e. high variance strategy has an initial frequency of 1/N).
c. The probability of invasion of the low variance strategy given the parametrs in 3a, with the low variance strategy has an intial frequency of 1/N.

**Figure 4**:
Individual based simulations for various deme sizes.
a. The fixation probability of the low variance strategy ($\sigma_1^2=9$, $\sigma_2^2=0.9$) is shown for various population sizes.
b.–d. These individual based simulations has the same parameters as for Figures 3a–c.

**Figure 5**:
"Effective population size" is plotted for different distributions of intrademic allele frequency (of either strategy) for a 10 deme metapopulation where the mean frequency is at 0.5. The populations chosen are {0.5, 0.5, 0.5, 0.5, 0.5, 0.5, 0.5, 0.5, 0.5, 0.5} (variance = 0), {0.1, 0.5, 0.5, 0.5, 0.5, 0.5, 0.5, 0.5, 0.5, 0.9} (variance = 0.0355), {0.1, 0.2, 0.3, 0.4, 0.5, 0.5, 0.6, 0.7, 0.8, 0.9} (variance = 0.067), {0.1, 0.1, 0.1, 0.1, 0.1, 0.9, 0.9, 0.9, 0.9, 0.9} (variance = 0.178), and {0,0,0,0,0,1,1,1,1,1} (variance = 0.277).
a. Migration rate = 0.01

b. Migration rate = 0.05

**Figure 6**:
Fixation probabilities for a range of migration rates for a system of D=10 demes, each with N=50 individuals. The mean fitness and variance parameters in parts a–c are the same as in Figure 3.

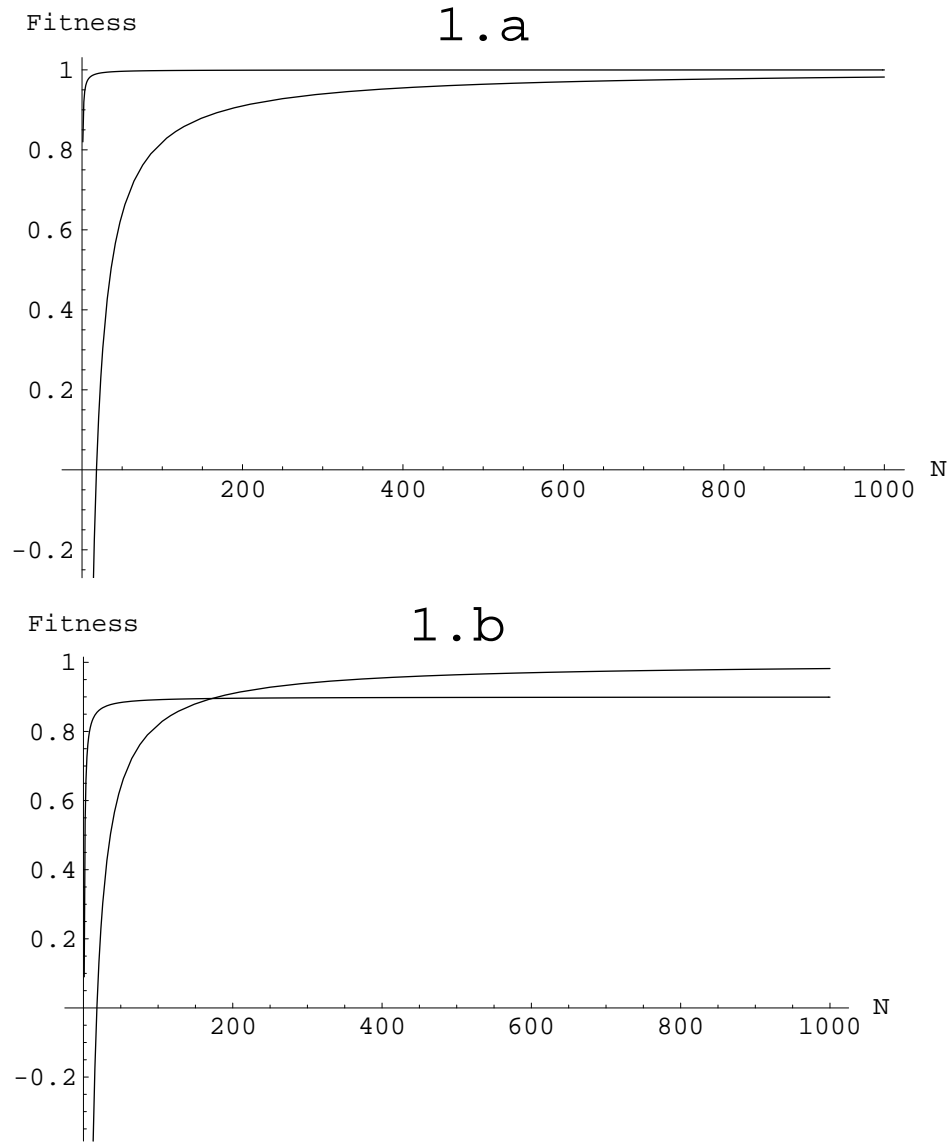

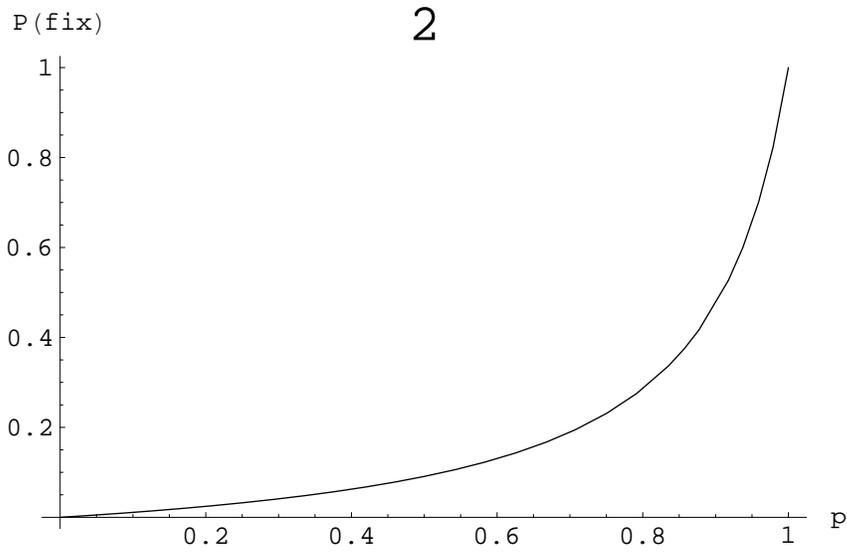

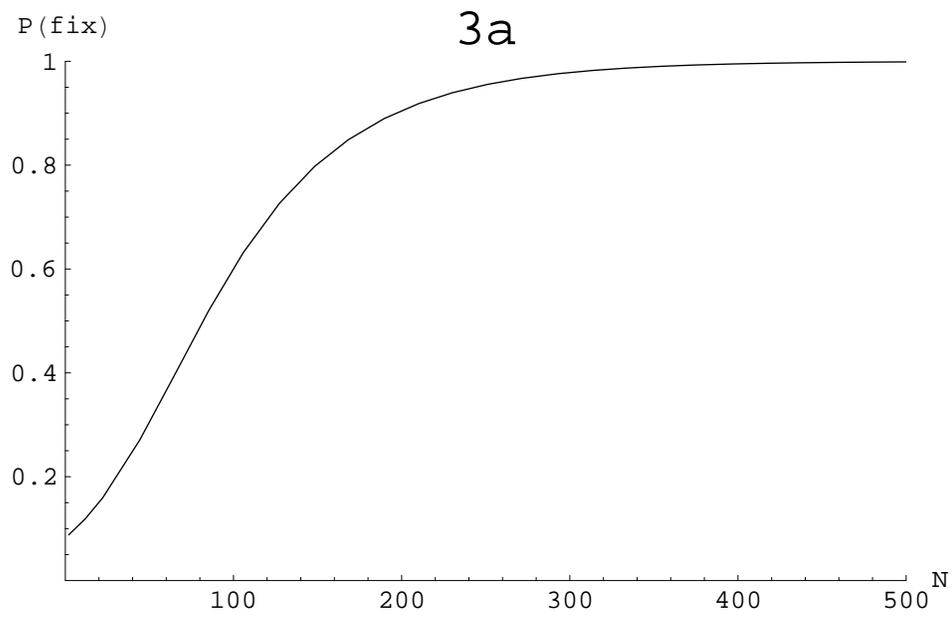

3a

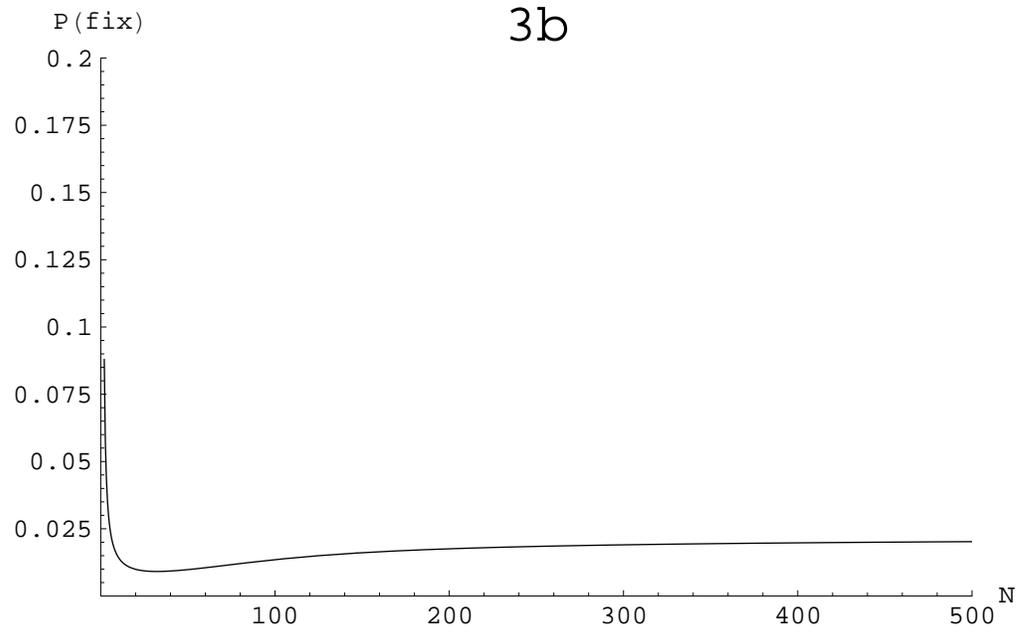

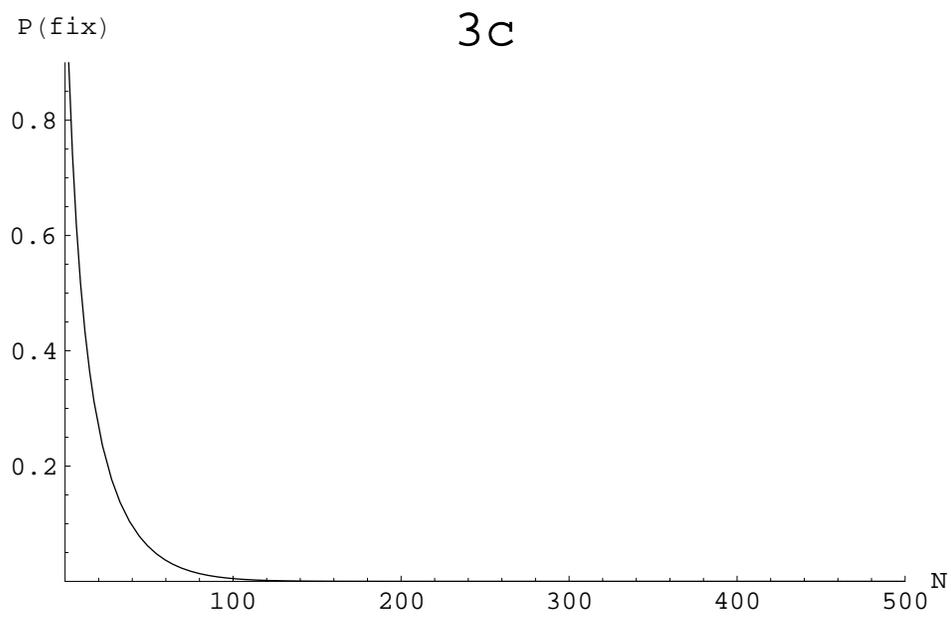

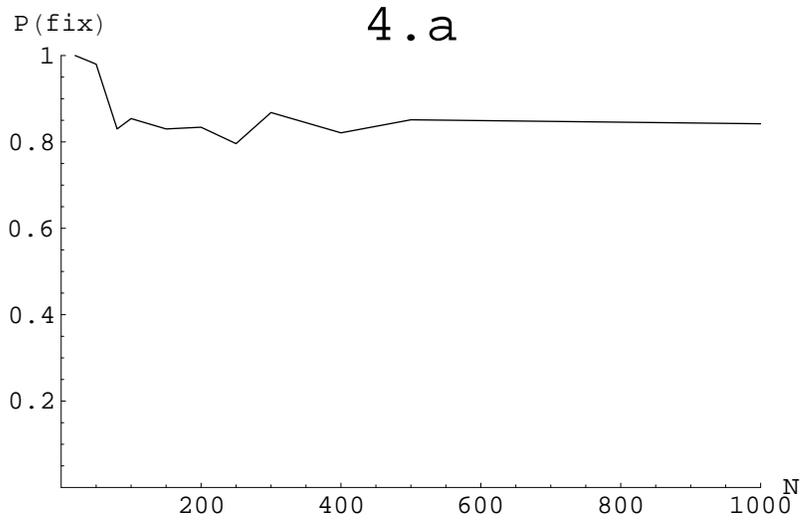

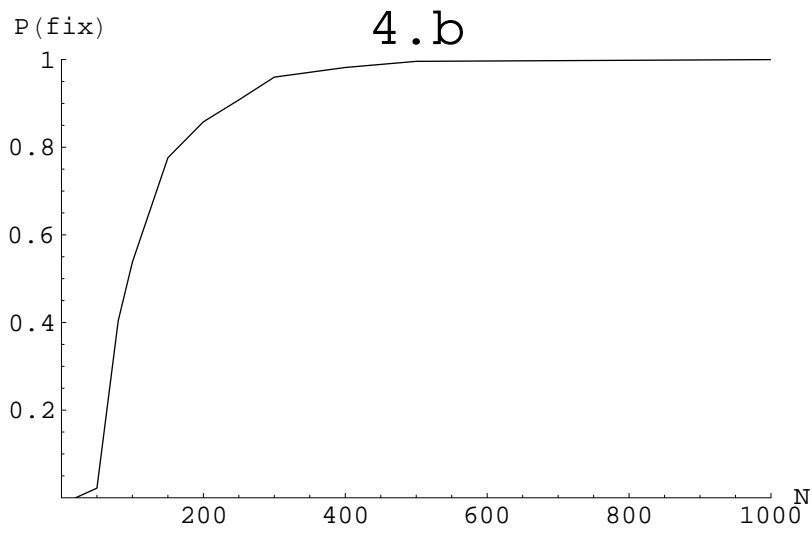

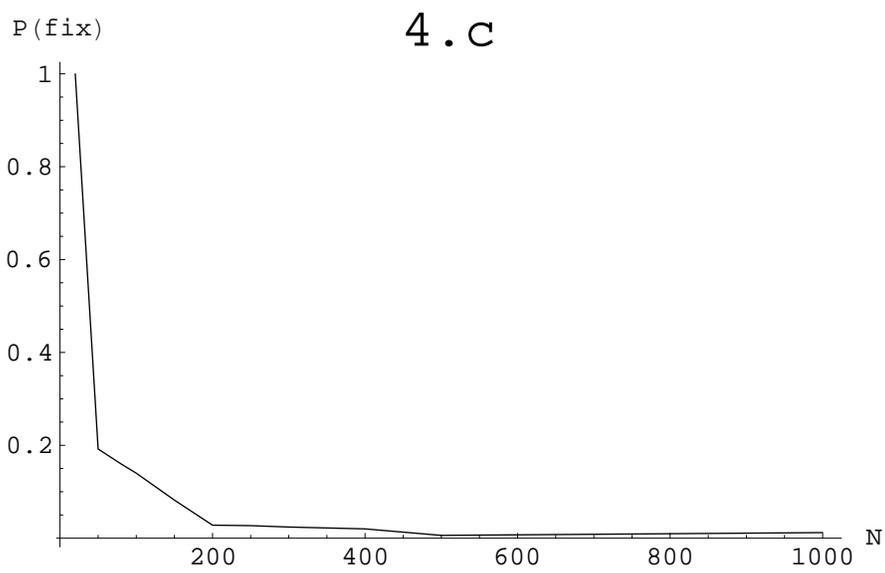

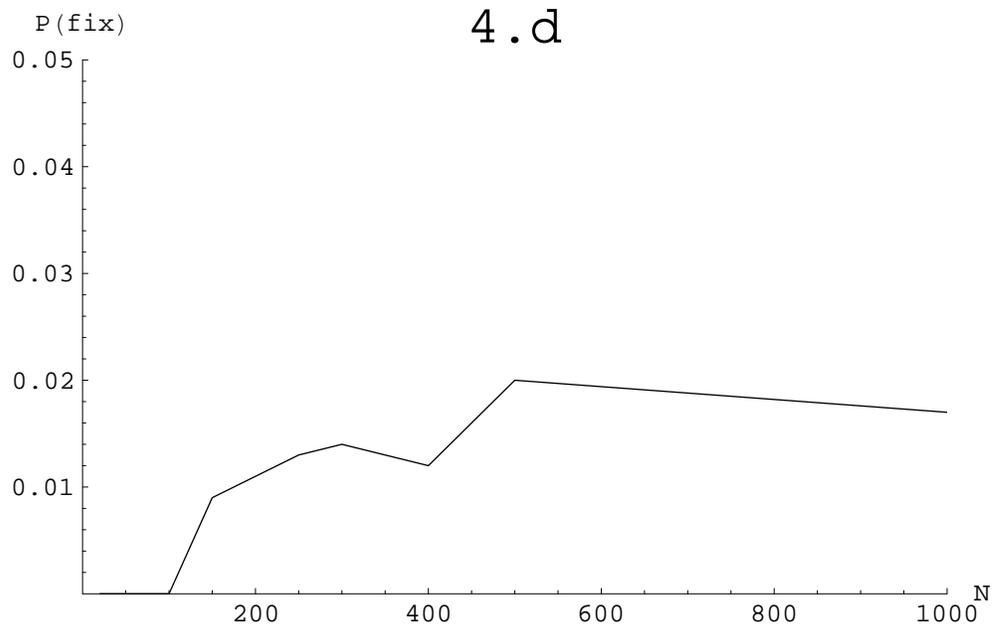

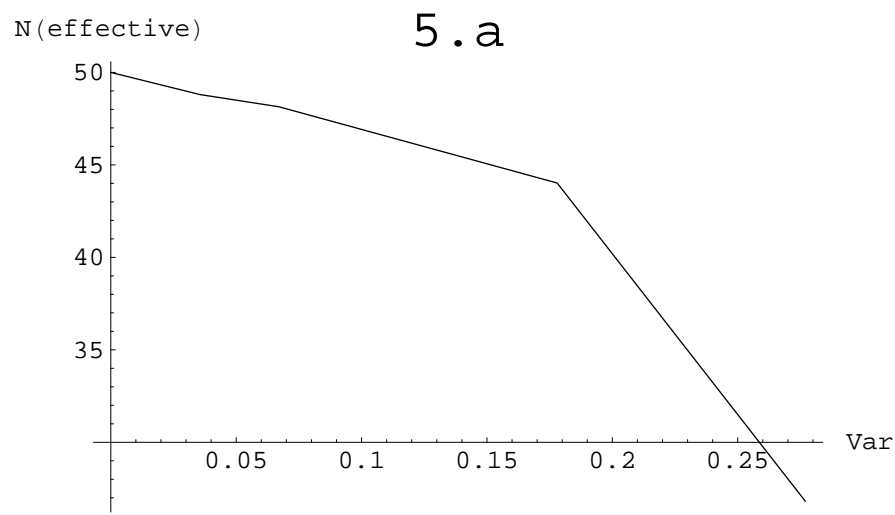

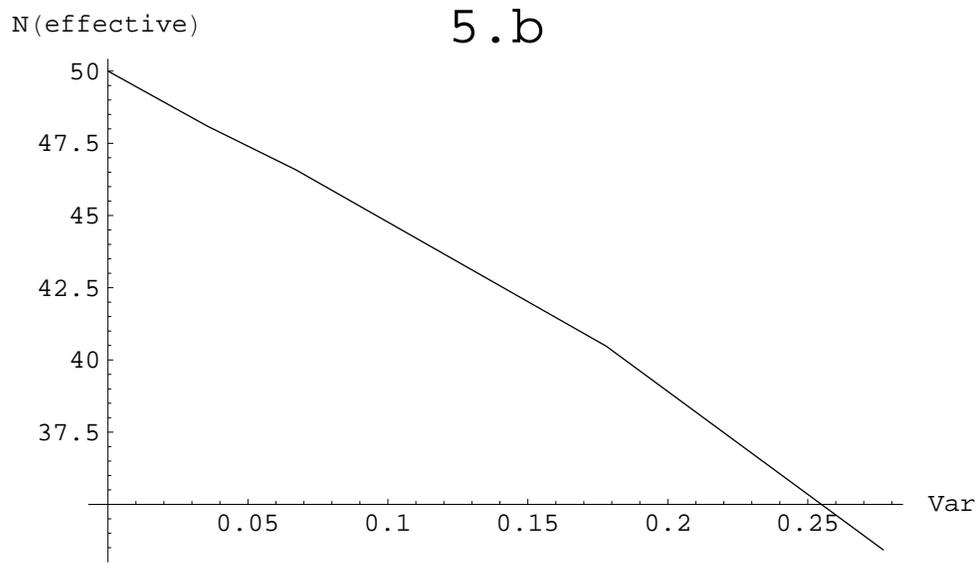

5.b

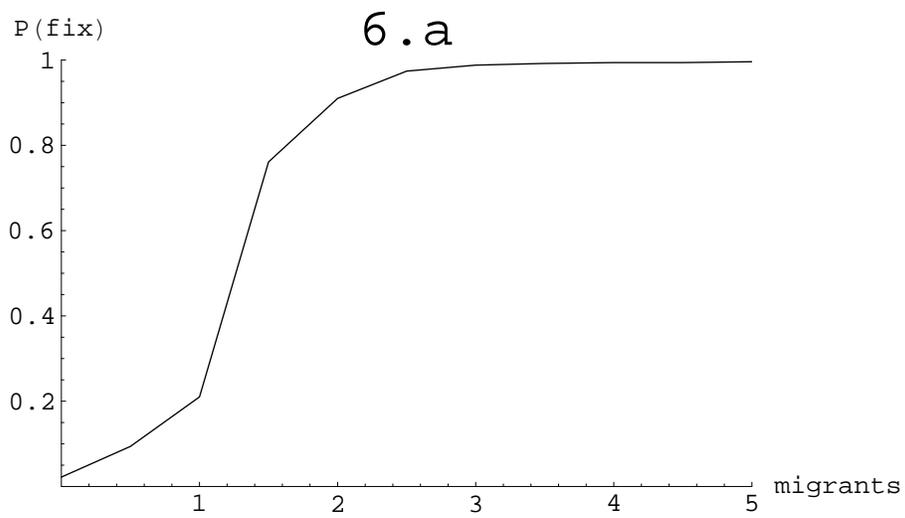

6.a

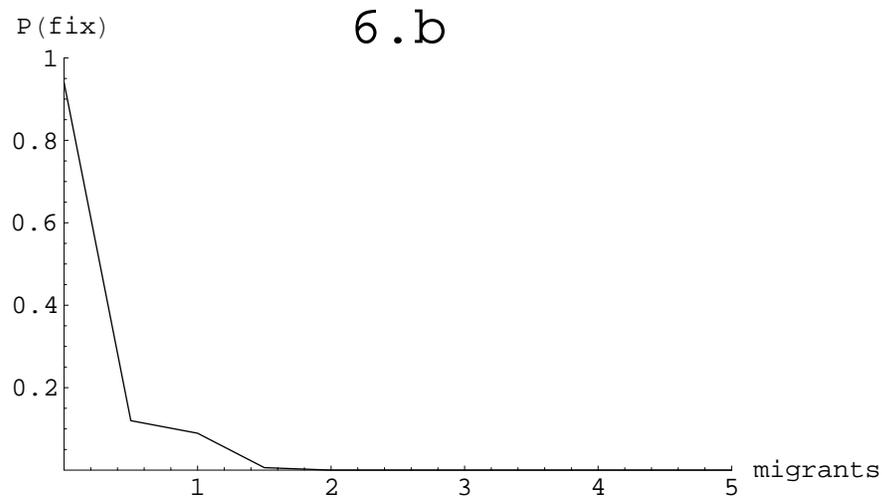

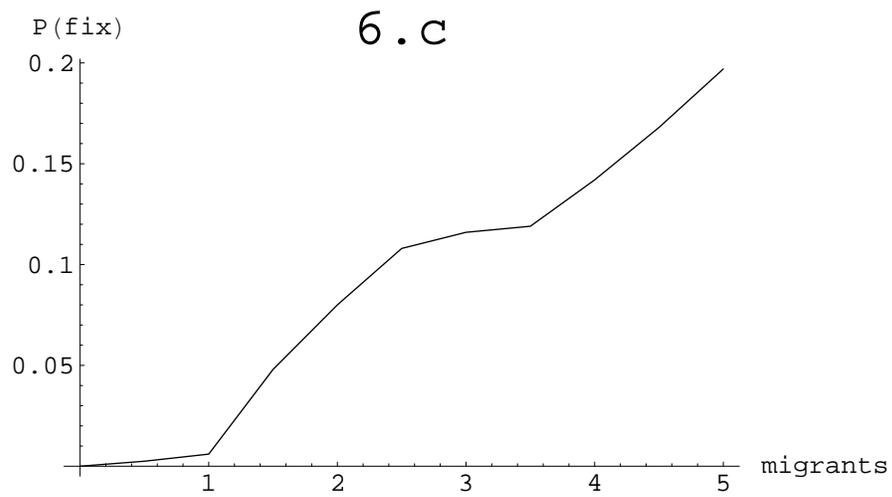